\documentclass[aps,prl,twocolumn,superscriptaddress,preprintnumbers,longbibliography]{revtex4-2}

\usepackage{amsmath}
\usepackage{graphicx}
\usepackage{epstopdf}
\usepackage{hyperref}
\usepackage{color} 

\DeclareMathOperator*{\sgn}{sgn}

\newcommand{\be}{\begin{equation}}
\newcommand{\ee}{\end{equation}}
\newcommand{\bea}{\begin{eqnarray}}
\newcommand{\eea}{\end{eqnarray}}
\newcommand{\nn}{\nonumber}
\newcommand{\bdm}{\begin{displaymath}}
\newcommand{\edm}{\end{displaymath}}

\newcommand{\dimreg}{\ensuremath{\epsilon_d}}

\newcommand{\mquad}{\ensuremath{\mathcal{M}_{Qg}}}
\newcommand{\pol}{\ensuremath{\varepsilon}}
\newcommand{\msggs}{\ensuremath{\mathcal{M}_{sggs}}}

\newcommand*\diff{\mathop{}\!\mathrm{d}}
\newcommand*\Diff[1]{\mathop{}\!\mathrm{d}^#1}
\newcommand{\intw}{\ensuremath{\int \frac{\diff \omega}{2 \pi}}}
\newcommand{\espect}[2]{\ensuremath{\frac{\diff E^{#1}_{#2}}{\diff \omega}}}

\usepackage{amssymb}
\usepackage[normalem]{ulem}

\newcommand{\aest}{\bgroup\markoverwith{\textcolor{red}{\rule[0.5ex]{2pt}{0.4pt}}}\ULon}

\begin{document}

\title{A tale of tails through generalized unitarity}

\author{Alex Edison}
%\email{alexander.edison@northwestern.edu}
\affiliation{Department of Physics and Astronomy, Northwestern University, Evanston, 
	Illinois 60208, USA}
\affiliation{Department of Physics and Astronomy, Uppsala University, 75108 Uppsala, Sweden}

\author{Mich\`ele Levi}
\email{levi@maths.ox.ac.uk}
\affiliation{Mathematical Institute, University of Oxford, Oxford OX2 6GG, United Kingdom}
\affiliation{Queen Mary University of London, London E1 4NS, United Kingdom}
\affiliation{Niels Bohr Institute, University of Copenhagen, 2100 Copenhagen, Denmark}

\date{\today}

\begin{abstract}
We introduce a novel framework to study high-order gravitational effects on a binary 
from the scattering of its emitted gravitational radiation.  
Here we focus on the radiation-reaction due to the background 
of the binary's gravitational potential, namely on the so-called tail effects, 
as the starting point to this type of scattering effects.  
We start from the effective field theory of a binary composite-particle. 
Through multi-loop and generalized-unitarity methods, we derive
the causal effective actions of the dynamical multipoles, 
the energy spectra, and the observable flux, due to these effects.  
We proceed through the third subleading such radiation-reaction effect, 
at the four-loop level and seventh order in post-Newtonian gravity, 
shedding new light on the higher-order effects, and pushing the state of the art.
\end{abstract}

%\keywords{,}
\preprint{UUITP-05/22}

\maketitle

\paragraph{Introduction.}

Since the first detection of gravitational waves (GWs) from a
black-hole binary merger \cite{Abbott:2016blz} by the Advanced LIGO
\cite{LIGOScientific:2014pky} and VIRGO \cite{VIRGO:2014yos}
collaboration, we have been rapidly shifting to a new era of
gravitational-wave astronomy. At present we already have a worldwide
network of second-generation ground-based GW experiments, including the
twin Advanced LIGO detectors in the US, Advanced Virgo in Europe
\cite{VIRGO:2014yos}, and the more recent KAGRA in Japan
\cite{KAGRA:2020tym}.  This network is planned to quickly expand, and
provide a steeply increasing influx of GW data of ever-higher quality
\cite{LIGOScientific:2018mvr,LIGOScientific:2020ibl,LIGOScientific:2021djp}.

These exciting developments on the experimental frontier go hand in
hand with a thrust in the theoretical frontier to push the program of
high-precision gravity. For present GW sources the inspiral phase, in
which typical velocities of the compact objects are non-relativistic,
has been studied analytically via the post-Newtonian (PN)
approximation of General Relativity \cite{Blanchet:2013haa}. PN
gravity forms the basis for theoretical generation of gravitational
waveforms, to be matched against measured data. This analysis is not
only probing new astrophysics and cosmology, but also new fundamental
physics, such as strong gravity and QCD in extreme conditions, which
cannot be produced on Earth \cite{LIGOScientific:2021sio}.

The surge in efforts to push the state of the art in PN gravity in the 
conservative sector has culminated at the fifth PN (5PN) order:
The point-mass potential was accomplished via a combination of traditional 
GR methods
\cite{Bini:2019nra,Bini:2020wpo,Bini:2020uiq}, and via the effective field
theory (EFT) approach \cite{Goldberger:2004jt,Blumlein:2020pyo}, and the
complete quadratic-in-spin interactions were accomplished via the EFT of 
spinning objects \cite{Levi:2015msa,Levi:2018nxp,Kim:2021rfj}. The completion 
of amplitude and phasing of radiation at the +4PN order (4PN orders 
beyond leading) is also currently underway \cite{Blanchet:2013haa}. 
Notably at these high orders there is an intricate class of
effects that come into play, which affect both the conservative and
radiative sectors. These effects are the scattering of
the binary's emitted radiation off its own background.

This scattering exerts radiation-reaction forces on the binary, and
contributes to the radiated energy-flux and to the binding energy of the
binary. While leading radiation that yields a radiation-reaction force
at the 2.5PN order contributes only to the radiated flux, the
subleading effect that first involves such scattering, the so-called
``tail'', enters at the 4PN order and already further affects the
conservative dynamics.  Such leading tail effects have been studied
for a few decades now using traditional GR methods
\cite{Blanchet:1987wq,Blanchet:1992br,Blanchet:1993ec,Blanchet:1993ng,Wiseman:1993aj},
which were extended to the next two subleading non-linear orders, the
so-called ``tail of tail'' (TT) and ``tail of tail of tail'' (TTT), in
\cite{Blanchet:1996wx,Blanchet:1997jj} and \cite{Marchand:2016vox},
respectively.

More recently, these effects have been studied via EFT methods, through
two different approaches. One approach involves the one-point function
of the stress-energy tensor as probed by an emitted on-shell radiation
graviton \cite{Goldberger:2009qd}, and proceeded through to the
TT non-linear order. The other approach, which was led by Galley
\cite{Galley:2008nss,Galley:2012hx,Galley:2014wla}, also provides the 
radiation-reaction forces on the binary. The latter was
applied to the leading radiation-reaction, namely without scattering
\cite{Galley:2009px,Galley:2012qs}, and proceeded only to the leading
tail effect \cite{Galley:2015kus}. Very recently, the 
subleading tail effects at 5PN order have been
approached in \cite{Foffa:2019eeb,Blanchet:2019rjs,Almeida:2021xwn}
and \cite{Blumlein:2021txe}.

In this letter we introduce a novel framework to study such
higher-order gravitational effects due to the scattering of radiation. 
%which we henceforth refer to as ``radiation-scattering''.
First, we note that at the radiation scale the scattered gravitons can go
on-shell, which naturally aligns with scattering-amplitudes
methods. This is unlike the situation in two-body
conservative interactions, where the exchanged gravitons
can never go on-shell.  Using amplitudes
methods at the orbital scale also alleviates the escalating
complications of standard EFT methods with Feynman calculus involving
the mixing of orbital and radiation modes.  

The main idea that we put
forward for the first time here is to think of the whole binary in analogy to
elementary massive particles with gravitons scattered off of them.  This
is inspired by long-observed analogies of gravitational interactions
of $l$-th mulitpole moments of a macroscopic object in effective
theories of gravity, to gravitational scattering amplitudes with
massive elementary particles of spin $l/2$, see
e.g.~\cite{Holstein:2008sx,Levi:2015msa}, and review in
\cite{Levi:2018nxp}.  Let us highlight though that the various
amplitudes-driven approaches that followed the latter, initiated in
\cite{Cachazo:2017jef,Bjerrum-Bohr:2018xdl,Cheung:2018wkq,Kosower:2018adc},
and recently reviewed in
\cite{Buonanno:2022pgc}, implement their methods on single
compact objects as elementary particles (and off-shell gravitions as noted), 
and thus have been tied to a treatment of the unbound problem of scattering 
two massive objects instead of the actual bound problem of the binary
inspiral.

In contrast, we advocate an entirely orthogonal approach. 
By treating the whole binary as elementary massive particles our
derivations lie directly in the binary inspiral problem and in PN
theory, and are consequently directly applicable to present and
planned GW experiments and measurements. In addition, a
connection of those previous approaches from the
unbound to the bound problem seems to become infeasible, even in 
restricted configurations, exactly when
radiation-reaction effects -- which we target in the novel formulation
in this letter -- show up \cite{Buonanno:2022pgc}.  Moreover, given
the current state of the art we need to push these effects to high
non-linear orders, which amounts to higher loops in QFT. 
Unlike previous works \cite{Buonanno:2022pgc}, in our
  present approach we do not invoke 
the propagation of quantum DOFs, which would in turn have to be
laboriously excised from the meaningful
classical contributions.  Rather we only work with classical
propagating DOFs, which keeps our formulation considerably lighter,
and thus more efficient for the problem at hand. 

In this letter we focus on scattering due to the binary's
gravitational potential, namely on tail effects, as the staring point
to tackle this generic type of radiation-scattering effects.  
We start from the EFT of a binary as composite particle, 
and use multi-loop integration \cite{Smirnov:2012gma,Smirnov:2019qkx} 
and generalized-unitarity methods
\cite{Bern:1994zx,Bern:1994cg,Britto:2004nc,Anastasiou:2006jv},
to set up a basis-unitarity inspired procedure to treat such effects,
assembling pure tree amplitudes generated by the public code
\texttt{IncreasingTrees} \cite{Edison:2020ehu} as building blocks. Since
time reversal no longer holds we invoke the closed time path (CTP)
formalism, which we extend to our new framework, and take a
radiation-reaction approach in order to uniquely capture the entirety
of effects -- on both conservative and dissipative sides. 
We
derive here the causal effective action of the dynamical multipoles, the
energy spectrum, and the observable flux due to these effects.  
 We proceed
through the third subleading effect, at the 4-loop level and 7PN
order, shedding new light on these higher-order effects, and pushing
the state of the art.

\paragraph{From a binary-particle EFT to generalized unitarity.} 

We start by recalling the effective action of a composite object
coupled to the gravitational field,
$g_{\mu\nu}\equiv\eta_{\mu\nu}+h_{\mu\nu}$, that reads
\cite{Goldberger:2009qd,Ross:2012fc,Levi:2018nxp}:
\begin{align} 
\label{seffcomp1}
S_{\text{eff(c)}}[g_{\mu\nu},y_c^\mu,e_{c\,A}^{\,\,\mu}] = 
& -\frac{1}{16\pi G}\int d^4x
\sqrt{g}\,R\left[g_{\mu\nu}\right] \nn \\
\,& +\, S_{\text{pp(c)}}
[g_{\mu\nu}(y_c),y_c^\mu,e_{c\,A}^{\,\,\mu}](\sigma_c), 
\end{align}
where $S_{\text{pp(c)}}$ is the worldline point-particle action 
of the composite particle with the form
\cite{Goldberger:2009qd,Levi:2010zu,Ross:2012fc,Levi:2018nxp}:
\begin{align}
\label{sppcomp}
S&_{\text{pp(c)}} [h_{\mu\nu},y_c^\mu,e_{c\,A}^{\,\,\mu}](t)= 
-\int dt \sqrt{g_{00}}\, \biggr[ E(t) \nn \\ 
+ &\frac{1}{2}\epsilon_{ijk}J^{k}(t)\left(\Omega_{\text{LF}}^{ij}
+\omega_\mu^{ij}u^\mu\right) - \sum_{l=2}^{\infty} 
\biggr(\frac{1}{l!}I^{L}(t)\nabla_{L-2}\mathcal{E}_{i_{l-1}i_{l}}%\right.&\nn\\
\nn \\
- &\frac{2l}{(l+1)!}J^{L}(t)\nabla_{L-2}\mathcal{B}_{i_{l-1}i_{l}}\biggr)\biggr], 
\end{align}
where here the worldline parameter is the time coordinate, $t$. 
$E$ here is the total energy of the composite object, and $I^L$ and
$J^{L}$ are definite-parity $SO(3)$ tensors, 
with the superscript $L$ for the indices $i_1\cdots i_l$ ($l\geq2$) 
in the Euclidean metric. 
They are coupled to the electric $\mathcal{E}$ and magnetic $\mathcal{B}$ components of the
Riemann tensor, respectively. 
For the present work we only need to consider the total energy, and the leading 
quadrupole moment, $I_{ij}$.

As the system is radiating and the symmetry of time reversal is
broken, the closed time path (CTP) formalism needs to be invoked
\cite{Calzetta:2008iqa,Galley:2008nss,Levi:2018nxp}, to integrate out
the gravitational field from \eqref{seffcomp1}. This yields a new
causal effective action of the binary multipoles, from which the
radiation-reaction forces and the energy spectrum of emitted
radiation can be derived. To switch onto the CTP formalism all degrees
of freedom (DOFs) are formally doubled, and the action is defined as:
\begin{equation}
  S_{\text{CTP}}[\{\}_1,\{\}_2]\equiv S[\{\}_1]-S^*[\{\}_2],
  \label{eq:ctp-split}
\end{equation}
where $\{\}$ denotes the set of all DOFs in the original action,
$S[\{\}]$.  For the doubled DOFs it is convenient to switch to the 
$\{+,-\}$ basis, which for classical fields entails the propagator matrix with 
the $\{+,-\}$ labels: $G_{++}=G_{--}=0$, $G_{+-}=G_{adv}$, $G_{-+}=G_{ret}$,
where the retarded and advanced propagators are given by 
\begin{equation}
G_{ret/adv}(x-x') = 
\int \frac{d^Dp}{(2\pi)^D} \frac{e^{-i p_\mu(x-x')^\mu}}
{(p_0 \pm i\epsilon)^2 - \vec{p}\, {}^2},
\label{eq:dret-def}
\end{equation}
namely the $+$ or $-$ $i \epsilon$ prescription for the retarded or
advanced propagator, respectively, and $D\equiv d+1$ with $d$ for 
the number of spatial dimensions. 

\begin{figure}[t]
\centering
\includegraphics[scale=0.5]{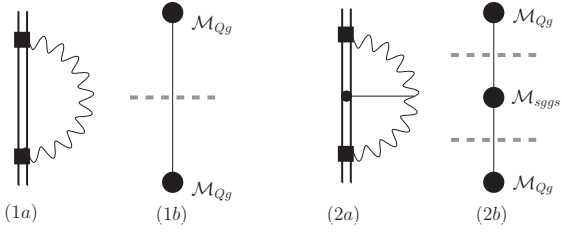}
\caption{Radiation-reaction and tail effects from the EFT and amplitudes perspectives. 
(1a), (2a): The Feynman graphs of maximal propagators, which contain all invariants of loop momenta.
The double line represents the worldline of the binary particle, the squares represent
quadrupole couplings, and the wiggly line represents the radiation graviton emitted.
The circle stands for the energy coupling, and the straight line for a potential graviton.
(1b), (2b): The amplitude cuts used to evaluate the radiation reaction and tail effects. 
The black circles represent tree amplitudes with massive particles, including a 4-particle 
amplitude of 2 scalars and 2 gravitons, sggs, and the solid lines cut by dashed lines stand 
for the graviton-state sewing of the cut.}
\label{uptotail}
\end{figure}

\begin{figure}[t]
\centering
\includegraphics[scale=0.5]{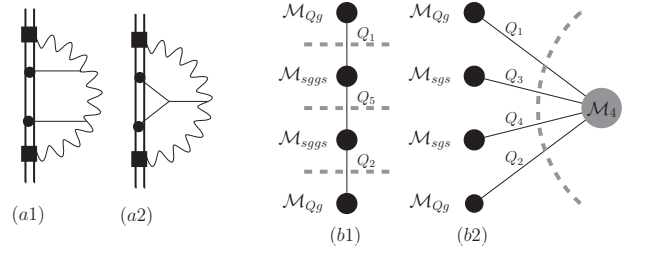}
\caption{The tail-of-tail effect from the Feynman and amplitudes
  perspectives. All notations are similar to figure \ref{uptotail},
  with the grey circle for a 4-graviton tree amplitude.  We similarly 
  only include the Feynman graphs with cubic vertices, which are used
  to determine all allowed propagators.}
\label{tailoftail}
\end{figure}

\begin{figure}[t]
\centering
\includegraphics[scale=0.5]{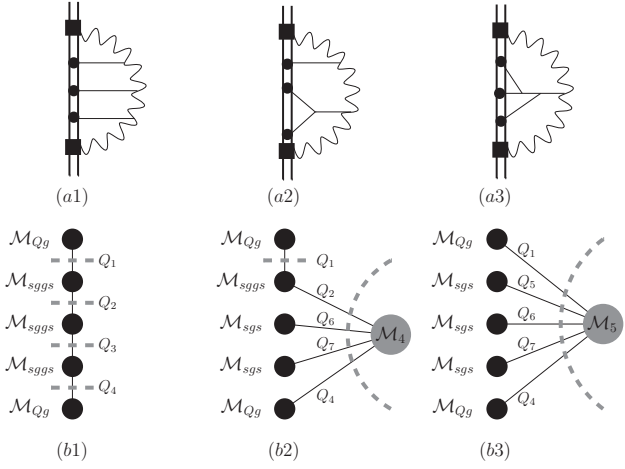}
\caption{The tail-of-tail-of-tail effect from the Feynman and
  amplitudes perspectives.  All notations and restrictions are similar
  to figures \ref{uptotail}, \ref{tailoftail}.  Graph a2 is also
  considered in its top-bottom mirror image, and graph b2 is evaluated
  as 2 cuts, which are swapped in a top-bottom mirror image. }
\label{tailoftailoftail}
\end{figure}

In the standard EFT approach the gravitational field is integrated out
using Feynman diagrammatic expansion. Figures \ref{uptotail}.1a,
\ref{uptotail}.2a, \ref{tailoftail}.a, and \ref{tailoftailoftail}.a show
example Feynman graphs that would need to be evaluated. Due to the
non-relativistic context, the integration is over 3-dimensional
spatial momenta, where the frequency of emitted radiation, $\omega$,
is regarded as the mass scale of such Euclidean propagators.
According to the generalized-unitarity paradigm, such Feynman
integration can be equivalently accounted for by writing the resulting
effective action as a linear combination:
\begin{equation}
  S_{\text{eff}}= \int
  \frac{d\omega}{2\pi} \sum_{i \in \text{MI}} c_{i}\mathcal{I}_{i}, 
\end{equation}
where $\{\mathcal{I}_{i}\}$ are a complete set of master integrals that span 
the integral family of the problem, and the coefficients ${c}_i$ are rational functions 
of the dimension $d$ and scales of the problem, which in our case is only the frequency. 
To fix the coefficients ${c}_i$ we will evaluate the cuts that span this complete set of integrals.

We illustrate how this new method works by treating
radiation-reaction and tail effects at increasing loop orders. First,
we approach radiation-reaction, depicted in figure \ref{uptotail}.  We
start by considering the Feynman graphs with the maximal number of
propagators that contain all possible invariants of loop
momenta. Radiation reaction has only one loop and thus one invariant,
captured by the single graph in figure \ref{uptotail}.1a. The integral
family at one-loop order is then simply:
\begin{align}
\label{eq:rr-family}
F^{(1)}(\lambda; \omega^2) & = 
\int \frac{\Diff{d}\ell_E}{(2 \pi)^{d}} \frac{1}{(-\ell_E^2 + \omega^2)^\lambda} \nn\\
& = R(d,\lambda,\omega^{2n(\lambda)}) F^{(1)}(1;\omega^2)\,, 
\end{align}
where for integer $\lambda$, $R$ is a rational function, with leading power of 
$\omega$ set by $\lambda$, 
and thus $F^{(1)}(1;\omega^2)$ is the master integral at one-loop order.
%Notice that at this point we still do not specify the causality
%prescription for the propagator.  Instead we keep the derivation
%generic, and defer the causality specification to later.  
We can then write the effective action as
\begin{equation}
S_{\text{RR}} = \intw c_{\text{RR}}(\omega) F^{(1)}(1;\omega^2) \,, 
\end{equation}
with $c_{\text{RR}}$ the coefficient to be fixed from unitarity cuts. 
Here we only need to evaluate one cut, as in figure \ref{uptotail}.1b.

Our cuts are assembled from tree amplitudes as building blocks, contracted via
graviton-state sewing, which inserts the relation:
\begin{align}
\label{eq:grav-sew}
\sum_{\text{states}} &\pol^{\mu \nu}_k \pol^{\alpha \beta *}_k 
\equiv \mathcal{P}^{\mu \nu; \alpha \beta}_k \nn\\
& = \frac{1}{2} \left( P_k^{\mu \alpha}P_k^{\nu \beta} + P_k^{\mu \beta}P_k^{\nu \alpha} - 
\frac{1}{D-2} P_k^{\mu \nu}P_k^{\alpha \beta} \right), 
\end{align}
in which
$ P^{\mu \nu}_k \equiv \eta^{\mu \nu} - \frac{k^\mu q^\nu + k^\nu
  q^\mu}{k \cdot q }, $ and $q$ is an arbitrary null reference
momentum, of which all dependence eventually cancels in any cut due to
gauge invariance \cite{Kosmopoulos:2020pcd}.
For %the amplitude that represents
the quadrupole coupling to the graviton, we make the following
definition:
\begin{align}
\label{eq:m-quad}
&\mquad  \equiv \lambda_Q J^{\mu \nu} \epsilon_{\mu\nu} 
\equiv \lambda_Q J^{\mu \nu} \pol_\mu \pol_\nu = - \lambda_Q I^{ab} \nn\\
& \hspace{5pt} \times \left(k_0 k_a \pol_0 \pol_b + k_0 k_b \pol_0 \pol_a - k_a k_b \pol_0 \pol_0 
- k_0 k_0 \pol_a \pol_b \right)\,,
\end{align}
with leading couplings only, and $\lambda_Q\equiv\sqrt{2\pi G_N}$.
The cut in figure \ref{uptotail}.1b is then assembled as:
\begin{align}
\mathcal{C}_{\text{RR}} & = \lambda_Q^2 J_1^{\mu \nu}\  
\mathcal{P}^{\mu \nu; \alpha \beta} J_2^{\alpha \beta} \Big|_{P_{\ell} 
	= \ell_E^2 - \omega^2 =0}\nn\\
&= \delta(\ell_E^2 - \omega^2)\lambda_Q^2 \left( J_1^{\mu \nu}J_2^{\mu \nu} 
- \frac{J_1^{\mu \mu} J_2^{\nu \nu}}{d-1} \right)\,, 
\end{align}
which evaluates to
\begin{align}
\mathcal{C}_{\text{RR}} = \delta(P_\ell) \lambda_Q^2 \frac{(d+1)(d-2)}{(d+2)(d-1)}  
\omega^4 \kappa_{ab}(\omega)\,, 
\end{align}
where $\kappa_{ab}(\omega) = I^{ij}_{a}(-\omega) I_{ij,b}(\omega)$
with $a,b\in\{+,-\}$, is the trace of the CTP quadrupole DOFs. 

The CTP effective action can then be written as  
\begin{align}
S_{\text{RR}} & = \frac{2 \pi G_N}{5} \intw \,\omega^4 \hspace{-5pt}
\sum_{a,b\in\{+,-\}}\hspace{-5pt}\kappa_{ab}(\omega) F^{(1)}(1_{ab}) \, ,
\end{align}
with the retarded and advanced propagators,
$ F^{(1)}(1_{-+}/1_{+-})\equiv F^{(1)}(1;(\omega \pm i\epsilon)^2)$, 
so that finally we obtain
\begin{equation}
\label{ctpactionRR}
S_{\text{RR}} = -i \frac{G_N}{5} \int_{-\infty}^\infty \frac{\diff{\omega}}{2 \pi} 
\,\omega^5 I_{-}^{ij}(-\omega) I_{+,ij}(\omega) \,, 
\end{equation}
in agreement with Galley et al.~in
\cite{Galley:2012qs,Galley:2015kus}, whose action is given in time
domain, and up to an overall sign discrepancy between the two
references \cite{Galley:2012qs,Galley:2015kus} -- we agree with the
latter.

Let us proceed to the tail effect that is captured by the single
Feynman graph depicted in figure \ref{uptotail}.2a. The
``integer-indexed'' integral family that contains the 3 invariants
constructed out of the 2 loop momenta reduces, using \texttt{FIRE6}
\cite{Smirnov:2019qkx}, to a master integral of only two propagators
for the two loops:
\begin{align}
F^{(2)}(1_X,1_Y,0) &= \int \frac{\Diff{d}\ell_{1}\Diff{d}\ell_{2}}{(2 \pi)^{2 d}} 
\frac{1}{(-\ell_{1}^2 + \omega^2_X)(-\ell_{2}^2 + \omega^2_Y)}\nn\\
&= F^{(1)}(1;\omega_X^2)F^{(1)}(1;\omega_Y^2)\,, 
\end{align}
where the entries in $F^{(2)}$ stand for exponents of the 3 denominators 
that span the generic integral family, and $X$, $Y$ label different possible 
$i\epsilon$ prescriptions. We can then write for the tail 
effective action:
\begin{equation}
\label{cutintailaction}
S_{\text{T}} = \intw c_{\text{T}}(\omega) F^{(2)}(1_X,1_Y,0) \,. 
\end{equation}

To assemble the cut that corresponds to this master integral and
determine $c_{\text{T}}$, we take a tree amplitude of 2 massive
scalars and 2 gravitons as a building block, corresponding to the
binary's energy $E$, coupling to two gravitons. 
This is where we use the analogy between the coupling of the binary's mass monopole to gravity 
and the gravitational scattering of massive scalar particles. 
The above amplitude can be extracted from \cite{Edison:2020ehu}, 
and since in the non-relativistic limit $|\vec{k}|,|\vec{p}| \ll m_s$, it is then expanded  
%\ace{Added a $_{COM}$ since this is the first appearance of $p$} 
in the large-mass limit as:
\begin{align}
\label{eq:sggs}
\msggs(m_s \to \infty) = & 
\frac{\lambda_g\lambda_E }{\omega_{k_2}^2} \frac{1}{2 (k_2^\mu k_{3,\mu})} 
\Big[ (k_2^\mu k_{3,\mu}) \pol_2^0 \pol_3^0 \nn\\
& + \omega_{k_2}( (\pol_3^\mu k_{2,\mu})\pol_2^0 - (\pol_2^\mu k_{3,\mu})\pol_3^0) \nn\\
& - \omega_{k_2}^2(\pol_2^\mu \pol_{3,\mu})\Big]^2 + \mathcal{O}(m_s^{-1}), 
\end{align}
where 2 and 3 label the two gravitons, and $\lambda_E$ is fixed from the 
3-particle tree amplitude of 2 massive scalars and a graviton,
%\begin{equation}
%\label{eq:m-mass}
$\mathcal{M}_{sgs} \equiv \lambda_E (p^{\mu}p^{\nu}/m_s^2) \pol^{\mu} \pol^{\nu}$, % \,, 
%\end{equation}
so that $\lambda_E\equiv-\sqrt{8\pi G_N} E$. The graviton self-coupling, 
$\lambda_g \equiv -\sqrt{32\pi G_N}$, is similarly fixed
from a 3-graviton amplitude \cite{Edison:2020ehu}.  With all the
ingredients in place we can assemble the cut, shown in figure \ref{uptotail}.2b, 
to fix the coefficient in \eqref{cutintailaction}:
\begin{align}
\mathcal{C}^{(2)}_{1,1,0} & = \sum_{\text{states}} \mathcal{M}_{Qg(-\omega)} \mathcal{M}_{sggs} 
\mathcal{M}_{Qg(\omega)} \Big|_{\substack{P_{\ell_{1}}=0,P_{\ell_{2}}=0 \\ m_s \to \infty}} \nn \\
& =  \lambda_Q^2 \, \delta(P_{\ell_{1}}) \delta(P_{\ell_{2}}) \nn\\
&	\times	J_{I(-\omega)}^{\mu \nu} P^{\mu \nu; \alpha \beta} 
\mathcal{M}_{sggs}^{\alpha \beta; \gamma \sigma} P^{\gamma \sigma; \rho \tau} 
J_{I(\omega)}^{\rho \tau} \Big|_{m_s \to \infty} \,.
\end{align}

Reducing the resulting integrals \cite{Smirnov:2019qkx}, and evaluating the master integrals 
with the appropriate CTP prescriptions, we finally find:
\begin{align}
\label{unrenctpactiontail}	
S_{\text{T}} = \frac{2}{5} G_N^2 E & \intw \,\omega^6 \kappa_{-+}(\omega)\nn\\ 
& \times\Bigg[\frac{1}{\dimreg} + \log \left(\frac{\omega^2}{\mu^2}\right) 
- i \pi \sgn (\omega) \Bigg],
\end{align}
with $\dimreg\equiv d-3$, and in agreement with equation (3.4) of
Galley et al., up to an overall sign discrepancy
\cite{Galley:2015kus}.  The coefficient of the
dimensional-regularization (DimReg) pole is \emph{even in $\omega$}.
When mapped from the $\{+,-\}$ to the $\{1,2\}$ CTP basis, terms that
are even in $\omega$ lead to a separated action for the quadrupoles of
the form eq.~\ref{eq:ctp-split}, and are thus conservative
\cite{Galley:2012hx,Galley:2014wla}.  Thus, as noted in
\cite{Galley:2015kus} this DimReg pole renormalizes the binding
energy.  We also absorb constant terms that are at the same $\dimreg$
order as the logarithmic term into the logarithm scale $\mu$.  We
apply similar implicit suppression to the following higher-order
results.

We proceed to the tail-of-tail (TT) effect, for which no effective
action has been previously derived.  We start again by considering the
Feynman graphs that contain the invariants from 3 loop momenta. 
In this case 2 graphs, shown in figure \ref{tailoftail}.a, suffice to 
contain all the invariants, which span the integral family at the 3-loop order:
\begin{align}
&F^{(3)}(\lambda_1,\lambda_2,\lambda_3,\lambda_4,\lambda_5,\lambda_6) \nn\\
&= \int \left(\prod_{i=1}^3 \frac{\Diff{d}\ell^i}{(2 \pi)^d} \right)
\frac{1}{Q_1^{\lambda_1}Q_2^{\lambda_2}Q_3^{\lambda_3} 
	Q_4^{\lambda_4}Q_5^{\lambda_5} Q_6^{\lambda_6}} \,, 
\end{align}
where the 6 invariants show up in the 6 denominators $\{Q_i\}$. 
For relevant integer values of $\lambda_i$ this integral is then reduced 
\cite{Smirnov:2019qkx}, and is found to be spanned by 2 master
integrals, so that we can write the effective action of the TT effect as:
\begin{align}
S_{\text{TT}} = \intw \Big[ & c_1(\omega) F^{(3)}(1,1,0,0,1,0) \nn \\
+ & c_2(\omega) F^{(3)}(1,1,1,1,0,0)\Big] \,, 
\end{align}
where again the 3-loop master integrals $F^{(3)}$ contain entries
for exponents of the 6 denominators, and we now suppress the
labels for various $i\epsilon$ prescriptions. 

The 2 cuts that correspond to these 2 master integrals are shown in
figure \ref{tailoftail}.b.  The first cut in \ref{tailoftail}.b1
is assembled from building blocks that we already used in lower loop orders: 
\begin{align}
\mathcal{C}^{(3)}_{1,1,0,0,1,0} = \lambda_Q^2 \, \delta(Q_1)\delta(Q_2)\delta(Q_5)& \nn\\ 
\times J_{I(-\omega)} P \mathcal{M}_{sggs,1} 
P\mathcal{M}_{sggs,2} & P J_{I(\omega)} \Big|_{m_s\to \infty} \,,
\end{align}
where the sewing indices were suppressed for readability, and
the resulting expression after evaluation is quite lengthy. 
The second cut in figure \ref{tailoftail}.b2 further requires the 4-graviton tree amplitude,
$\mathcal{M}_4$, taken from \cite{Edison:2020ehu}, to which no special
kinematics should be applied for our context. 
This cut is assembled as follows:
\begin{align}
\label{eq:tt-bc-ms}
&\mathcal{C}^{(3)}_{1,1,1,1,0,0} = 
\lambda_Q^2 \, \delta(Q_1)\delta(Q_2) \delta(Q_3) \delta(Q_4) &\nn\\
&\times (J_{I(-\omega)} P)(\mathcal{M}_{sgs,1} P) 
\mathcal{M}_4^{\text{tree}}  (P \mathcal{M}_{sgs,2})(P J_{I(\omega)})\,, 
\end{align}
where again we suppress the contraction indices for readability.
Plugging in the values of cuts and the appropriate CTP
prescriptions, we finally find the CTP effective action of the TT
effect:
\begin{align}
\label{unrenctpactionTT}
S_{\text{TT}} = \frac{107}{175} \, G_N^3 E^2 & \intw \,
  \omega^7 \kappa_{-+}(\omega) \nn\\
  \times \Bigg[ \pi \sgn(\omega) & + i \left[\frac{2}{3\,\dimreg} 
  + \log\left(\frac{\omega^2}{\mu_1^2}\right)\right] \Bigg] \,.
\end{align}
Unlike in the tail effective action, the DimReg pole is now
non-conservative, as its coefficient is \emph{odd in $\omega$} leading
to a  CTP action that cannot be separated as in eq.~\ref{eq:ctp-split}
\cite{Galley:2012hx,Galley:2014wla}.  Thus, it must be removed prior
to extracting dissipative observables from the action.  The most
straightforward method of removal is to introduce a renormalized
coupling to the quadrupole, similar to \cite{Goldberger:2009qd} (see
Appendix).

Building on the procedure presented at lower loop orders, we briefly
outline the derivation for the tail-of-tail-of-tail (TTT) effect, which
proceeds along similar lines. There are 4 Feynman graphs, shown in figure
\ref{tailoftailoftail}.a, that span the integral family at the 4-loop
order with 10 generic denominators. The relevant integrals are then
reduced to 4 master integrals \cite{Smirnov:2019qkx}, so that the
effective action of the TTT effect can be written as:
\begin{align}
S_{\text{TTT}} & = \intw \Big[ c_1\, F^{(4)}_{1,1,1,1,0,0,0,0,0,0}  
+ c_2\, F^{(4)}_{1,0,0,1,1,1,1,0,0,0} \nn\\
+ \big( & c_3\, F^{(4)}_{1,0,1,1,1,1,0,0,0,0} + c_4\, F^{(4)}_{1,1,0,1,0,1,1,0,0,0}\big) \Big] \,, 
\end{align}
where the entries in $F^{(4)}$ are for exponents of the 10 denominators, 
and we suppress labels for various $i\epsilon$ prescriptions and dependence in $\omega$ of the
coefficients $c_i$. %to be fixed from evaluating the corresponding cuts. 

The 4 cuts that correspond to the 4 master integrals are shown in
figure \ref{tailoftailoftail}.b, where the cut
$\mathcal{C}^{(4)}_{1,0,0,1,1,1,1,0,0,0}$ in \ref{tailoftailoftail}.b3,
further requires the 5-graviton tree amplitude, $\mathcal{M}_5$, taken
from \cite{Edison:2020ehu}.  The cuts are then assembled as in the
previous cases, and each is thousands of terms long to begin with.
Substituting in the values of cuts and the proper CTP
prescriptions, we arrive at the CTP effective action of the TTT
effect:
\begin{align}
\label{unrenctpactionTTT}
S_{\text{TTT}} =
- \frac{4}{525} G_N^4 E^3 \intw  \, \omega^8 \kappa_{-+}(\omega) & \nn\\
\times\Bigg[ \frac{107}{2\dimreg^2}
+ \frac{107}{\dimreg} \log \left(\frac{\omega^2}{\mu_2^2}\right) 
+ 107\log^2 & \left(\frac{\omega^2}{\mu_2^2}\right) \nn\\
+ \frac{20707426967}{60399360}  - \frac{3103}{4}\,\zeta_2 - 420 \, & \zeta_3  \nn\\
- i \pi \sgn(\omega) \Bigg[\frac{107}{\dimreg} 
+ 214 \log \Big(&\frac{\omega^2}{\mu_2^2}\Big) \Bigg]\,\Bigg]\,,
\end{align}
where $\zeta_2\equiv \pi^2/6$ and $\zeta_3$ is Ap\'ery's constant. 
The DimReg poles now appear in both the conservative
and dissipative parts. The non-conservative pole can be removed by
renormalizing quadrupoles in the tail action, using exactly the
same renormalization scheme as in TT.

\paragraph{From CTP Effective Actions to Spectra and Fluxes.} 

It is useful to have the CTP effective actions in order to obtain the
related radiation-reaction forces by varying with respect to the CTP
DOFs, $\{q_{\pm}^i\}$, and then taking the physical limit,
$q_+^i \to q^i$ and $q_-^i \to 0$. We defer a
discussion of the conservative sector for future work.

We can extract the energy spectrum in the CTP formalism by
starting from the generalized Noether theorem \cite{Galley:2014wla},
which tells us that in the time domain:
\begin{equation}
	\frac{\diff E}{\diff t} = - \frac{\partial L}{\partial t} 
	+ \dot{q}^i \left[ \frac{\partial K}{\partial q_-^i} \right]_{\text{PL}} 
	+ \ddot{q}^i \left[ \frac{\partial K}{\partial \dot{q}_-^i} \right]_{\text{PL}} 
	+ \ldots \,,
	\label{eq:ctp-gen-flux}
\end{equation}
where $L$ is the conservative potential of \emph{one} of the time
histories, $K$ is the non-conservative potential, $\{q^i\}$ are the
generalized coordinate variables or DOFs, and PL denotes the physical
limit as noted, $q_+ \to q$ and $q_- \to 0$. We then work out \eqref{eq:ctp-gen-flux}
with the CTP quadrupoles as our generalized DOFs, and if we then
integrate over $t$, we arrive at
\begin{equation}
	\int \diff t \frac{\diff E}{\diff t} = \Delta E = \int \diff \omega \espect{}{}, 
\end{equation}
where on the right-hand side we have the energy spectrum that we want.

Applying our generic derivation to the tail actions is then straightforward. 
First, we obtain the energy spectrum of radiation reaction as:
\begin{align}
	\int_{0}^\infty \diff{\omega}\espect{}{\text{RR}} = 
	-\frac{G_N}{5 \pi} \int_{0}^\infty \diff{\omega}\,
	\omega^6 \kappa(\omega)\,, 
\end{align}
where now $\kappa(\omega)\equiv I^{ij}(-\omega)I_{ij}(\omega)$.
Similarly, we obtain the following power spectra:
\begin{align}
	\label{tailspect}
	\espect{}{\text{T}} & = - \frac{2}{5} G_N^2 E \,\omega^7 \kappa(\omega) \,, \\ 
	\label{eq:dele-tt-eval}
	\espect{}{\text{TT}} & = \frac{428\,\,}{525\pi} G_N^3 E^2 \,
	\omega^8 \log(\omega/\mu_1)\kappa(\omega)  \,,  \\
	\espect{}{\text{TTT}} & = \frac{856}{525} G_N^4 E^3 \,
	\omega^9 \log(\omega/\mu_2) \kappa(\omega) \,, 
	\label{eq:ttt-spect}
\end{align}
for the tail, TT and TTT effects, respectively. 
\eqref{tailspect} and \eqref{eq:dele-tt-eval} are in agreement with \cite{Bini:2021qvf}, 
and \eqref{eq:ttt-spect} is new.

As a final check, we can also specialize to a circular orbit with orbital
frequency $\Omega$ to get the energy flux in terms of the symmetric
mass ratio $\nu$, and the PN parameter $x=(\Omega G_N E)^{2/3}$. We
then obtain:
\begin{align}
	P_{\text{RR}}^{\text{circ}} = - \frac{32}{5G_N } \nu^2 x^5,  \quad %\\ 
	P_{\text{T}}^{\text{circ}} = -\frac{128 \pi}{5 G_N} \nu^2 x^{13/2} \,.
\end{align}
For the TT and TTT we present the 
%scale-invariant 
non-analytic contributions:
\begin{align}
	P_{\text{TT}}^{\text{circ}}  = \frac{27392}{175 G_N} \nu^2 x^8 \ln x, \,\,\, %\\
	P_{\text{TTT}}^{\text{circ}} = \frac{109568 \pi }{175 G_N} \nu^2 x^{19/2} \ln x \,.
\end{align}
These results for the flux from a circular orbit are in complete agreement with
\cite{Tanaka:1996lfd, Fujita:2012cm, Blanchet:1997jj,
	Blanchet:2013haa, Marchand:2016vox}.

\paragraph{Future prospects of the new unitarity framework.} 

In this letter we introduced a novel framework to tackle higher-order
gravitational effects due to scattering of the binary's emitted
radiation from its own gravitational background.  Within this
framework we derive the causal effective actions of the dynamical
multipoles, that encapsulate all conservative and dissipative physics,
including those of the TT and TTT, that were never previously
derived. We derive dissipative observables: first the
generic energy spectra, and then the observed
circular-orbit flux due to these effects. We
find complete agreement with available results obtained via
traditional GR and standard EFT methods.  One can also derive the
conservative dynamics from our actions, e.g.~EOMs and binding
energies. Given the current state of the art we set out to establish a
framework which is able to push through these effects to higher PN
orders. Here we proceeded through the third subleading such
radiation-reaction effect, which corresponds to the 4-loop level and
the 7PN order. This is shedding new light on these higher-order
effects, and pushing the state of the art.

Our novel framework utilizes multi-loop and generalized-unitarity
methods to set up an amplitudes-like computation which captures such
radiation-scattering effects with high efficiency. In this letter we
demonstrated that the new approach is already competitive with
traditional GR methods, and even outpaces standard EFT methods, which
become intractable already at subleading tail effects. Our framework
constitutes the first direct application of modern amplitude methods
to the binary inspiral problem and thus to present and planned GW
measurements, in PN theory. Obvious extensions of this framework
include subleading PN orders of the non-linear effects, and scattering
of subleading radiation from generic multipole sources off background
generated by generic multipole sources.  Both entail tree amplitudes
with massive particles of any spin, namely also of higher spins.  As
noted the framework presented here should be straightforward to apply 
to the conservative as well as the radiative sector. 
We leave all these developments for future work.

\begin{acknowledgments}
	
\paragraph{Acknowledgments.}
We thank Donato Bini and Luc Blanchet for pleasant discussions.
AE is supported in part by the Knut and Alice Wallenberg Foundation
under KAW 2018.0116, by Northwestern University via the Amplitudes and
Insight Group, Department of Physics and Astronomy, and Weinberg
College of Arts and Sciences, and by the US Department of Energy under
contract DE-SC0015910.
ML received funding from the European Union's Horizon
2020 %research and innovation programme
under the Marie Sk{\l}odowska-Curie grant 847523, and has been
supported by the Science and Technology Facilities Council (STFC)
Rutherford Grant ST/V003895 \textit{``Harnessing QFT for Gravity''},
and by the Mathematical Institute University of Oxford.
\end{acknowledgments}

\appendix

\section*{Appendix: Renormalizing Higher-Order Tails}

The CTP effective actions in eqs.~\eqref{unrenctpactiontail}, \eqref{unrenctpactionTT}, 
\eqref{unrenctpactionTTT} contain DimReg poles, and go through a renormalization. 
As we illustrate below, there is an interplay among lower-order DimReg zeros and higher-order 
DimReg poles, similar to that in purely conservative effective potentials as of the N$^3$LO 
sectors. 
%as of the 3PN order (which starts going beyond the PN accuracy of the LO radiation-reaction!). 
The renormalization we apply is essentially similar to that in \cite{Goldberger:2009qd}, where the 
quadrupole moment gets renormalized and displays an RG flow as 
a Wilson coefficient of the EFT at the radiation scale.
Here we shall demonstrate the renormalization needed for the extraction of dissipative 
physics, which was discussed in the above. 

First, we note that the CTP effective action of radiation-reaction actually contains a piece 
proportional to a simple DimReg zero beyond the leading expression presented in 
eq.~\eqref{ctpactionRR}:
\begin{align}
\label{dimregzeroRR}
\Delta S_{\text{RR}}\Big|_{\dimreg^1}=&\dimreg \frac{G_N}{20} \int \frac{d\omega}{2\pi} \omega^5 
\Big[ -\pi \sgn\omega \big(\kappa_{+-}(\omega) + \kappa_{-+}(\omega) \big) \nn\\
&+i \bigg(\frac{9}{10}-\gamma_E+ \log\pi-\log\frac{\omega^2}{\mu_0^2}\bigg)\nn\\
&\big(\kappa_{+-}(\omega) - \kappa_{-+}(\omega) \big)\Big].
\end{align}
In the effective
action of the tail, eq.~\eqref{unrenctpactiontail}, the DimReg pole
(and corresponding logarithm) is purely in the conservative part of
the effective action, so it does not affect dissipative observables.

The first dissipative DimReg pole occurs in the TT effective action,
eq.~\eqref{unrenctpactionTT}.
Following textbook renormalization procedures, (and Ref.~\cite{Goldberger:2009qd}'s 
application in a similar context)
we introduce a renormalized coupling to the quadrupoles:
\begin{equation}
\label{rgflow}
\kappa_{ij} \to \overline{\kappa}_{ij} \equiv 
\kappa_{ij} \Bigl( 1+\frac{214}{105} \dimreg^{-1} \, G_N^2E^2\omega^2 \Bigr). 
\end{equation}
With this substitution in eq.~\eqref{ctpactionRR} we find that
\begin{equation}
\overline{S}_{TT} \equiv \left(S_{RR} + S_{T} +  S_{TT}\right) \Big|_{\kappa_{ij} 
	\to \overline{\kappa}_{ij}}
\end{equation}
is free of dissipative DimReg poles through $\mathcal{O}(G_N^3)$.
Extracting the $\mathcal{O}(G_n^3)$ contribution from
$\overline{S}_{TT}$ defines the
renormalized TT effective action:
\begin{align}
  S_{\text{TT}}^{\text{Ren}} %&\equiv \overline{S}_{TT}\Big|_{G_N^3} \notag \\
  &= 
\frac{G_N^3E^2}{10} \int \frac{d\omega}{2\pi} \omega^7
\kappa_{-+}(\omega)
\Bigg[\frac{428}{105}\Big[\pi\text{sgn}\omega + \nn\\
&i\Big(\gamma_E-\text{log}\pi+\text{log}\frac{\omega^2}{\mu_0^2}\Big)\Big]
  -i\Bigg(\frac{634913}{22050}+16\zeta_2 \Bigg)\Bigg].
\end{align}
While the TTT effective action, eq. \eqref{unrenctpactionTTT},
contains higher-order DimReg poles, the dissipative part only
contains a simple pole.  
As such, the same renormalization in eq.~\eqref{rgflow}
applied to the $\mathcal{O}(\dimreg^1)$ part of the tail is
sufficient to remove the dissipative TTT pole and obtain the
renormalized action $\overline{S}_{TTT}$.  
We defer a discussion of the full renormalization including the conservative sector 
to future work.

With renormalized couplings, we also expect an RG flow of the
quadrupoles (or equivalently $\kappa_{ij}$).  The flow equation can
be found by allowing $\overline{\kappa}$ to depend on the log scale
$\mu_0$ then demanding that $\overline{S}_{TTT}$ does not depend on $\mu_0$.  
Doing so, we find
\begin{equation}
\label{eq:rgflow}
\frac{d}{d \log \mu} \overline{\kappa} = - \frac{428}{105}(G_N \omega E)^2 \overline{\kappa},
\end{equation}
in exact agreement with \cite{Goldberger:2009qd} (noting that
  $\frac{d}{d \mu} \kappa \sim 2 \frac{d}{d \mu} I_{ij}$).

\bibliography{gwbibtex}

\end{document}